\documentclass[unnumsec,webpdf,contemporary,large]{oup-authoring-template}%

\graphicspath{{Fig/}}

\theoremstyle{thmstyleone}%
\theoremstyle{thmstyletwo}%
\theoremstyle{thmstylethree}%

\newcommand{\ech}{\color{black}\rm}

\begin{document}

\journaltitle{ }
\DOI{DOI HERE}
\copyrightyear{2025}
\pubyear{2025}
\access{Advance Access Publication Date: Day Month Year}
\appnotes{Original Paper}

\firstpage{1}


\title[TopSpace]{TopSpace: spatial topic modeling for unsupervised discovery of multicellular spatial tissue structures in multiplex imaging}

\author[1,$\ast$]{Junsouk Choi}
\author[2]{Jian Kang}
\author[2]{Veerabhadran Baladandayuthapani}

\authormark{J. Choi et al.}

\address[1]{\orgdiv{Department of Statistics}, \orgname{Korea University}, \orgaddress{\street{Seoul}, \country{Korea}}}
\address[2]{\orgdiv{Department of Biostatistics}, \orgname{University of Michigan}, \orgaddress{\street{Ann Arbor}, \state{MI}, \country{USA}}}

\corresp[$\ast$]{Corresponding authors. \href{email:junsoukchoi@korea.ac.kr}{junsoukchoi@korea.ac.kr}}

\received{Date}{0}{Year}
\revised{Date}{0}{Year}
\accepted{Date}{0}{Year}



\abstract{\textbf{Motivation:} Understanding the spatial architecture of tissues is essential for decoding the complex interactions within cellular ecosystems and their implications for disease pathology and clinical outcomes. Recent advances in multiplex imaging technologies have enabled high-resolution profiling of cellular phenotypes and their spatial distributions, revealing critical roles of tissue structures such as tertiary lymphoid structures (TLSs) in shaping immune responses and influencing disease progression. However, existing methods for analyzing spatial tissue structures often rely on hard clustering or adjacency-based spatial models, which are limited in capturing the nuanced and overlapping nature of cellular communities. To address these challenges, we develop a novel spatial topic modeling framework for the unsupervised discovery of spatial tissue structures in multiplex imaging data. \\
\textbf{Results:} We propose TopSpace, a novel Bayesian spatial topic model that integrates Gaussian processes into latent Dirichlet allocation---a modern machine learning topic modeling approach in natural language processing---to flexibly model spatial dependencies in tissue microenvironments. By leveraging the Bayesian framework, TopSpace supports multicellular mixed-membership clustering and offers key inferential advantages, including robust uncertainty quantification and data-driven determination of the number of multicellular microenvironments. We demonstrate the utility of TopSpace through simulations and a case study on non-small cell lung cancer (NSCLC) data. Simulations show that TopSpace accurately recovers latent tissue microenvironments and spatial clustering patterns, outperforming existing methods in scenarios with varying spatial dependencies. Applied to NSCLC data, TopSpace successfully identifies TLS and captures their spatial probability distribution, which strongly correlates with patient survival outcomes. \\
\textbf{Availability:} The TopSpace R package is available at \url{https://github.com/junsoukchoi/TopSpace.git}.
}


\maketitle

\section{Introduction}\label{sec:intro}

The functioning of biological systems in human tissues such as tumors depends on complex interactions among various cell types within  complex, structured, and dynamic cellular ecosystems \citep{Okabe2016-tw}. In instances of disease, malfunction often spans cells  of  multiple  types  and  entails alterations  in  the  composition, structure and organization of tissues.  Therefore, understanding higher-level patterns and spatial structures of tissues is fundamental to understanding disease pathology and predicting clinical outcomes \citep{Palla2022-se}. Recent development of spatial multiplex imaging technologies such as cyclic immunoflourescene (CyCIF, \citealp{Lin2015-oe}), CO-Dectection by indEXing (CODEX, \citealp{Goltsev2018-ei}), multiplex immunohistochemistry (mIHC, \citealp{Tsujikawa2017-dd}), and multiplex ion beam imaging (MIBI, \citealp{Angelo2014-xr}) allows for measuring the expression of multiple markers at single-cell resolution while preserving spatial information of cells. This enables direct observation of cellular phenotypes along with their spatial distributions and interactions across the tissue domain. This provides a highly resolved view of cellular heterogeneity in tissues and have already generated insights into the spatial organization of diverse multicellular organisms such as developmental brain tissues \citep{Moffitt2018-ij} and tumor microenvironments \citep{Stahl2016-wu}.

Recent studies have demonstrated the critical role of spatial tissue structures in shaping immune responses and influencing disease outcomes, especially in cancer. One prominent example is the study of tertiary lymphoid structures (TLS), which are organized aggregates of immune cells that form within tumors \citep{Sautes-Fridman2016-tj}. The presence of TLS has been associated with improved prognosis and better responses to immunotherapies in various cancers, as they facilitate local immune responses against tumor cells \citep{Fridman2017-zr}. Multiplex tissue imaging allows for detailed characterization of the cellular composition and spatial organization of TLS, providing insights into their role in anti-tumor immunity. Another example is the spatial distribution of cytotoxic CD8+ T cells within the tumor microenvironment. Studies have shown that not just the abundance but the localization of these immune cells in proximity to tumor cells correlates with patient outcomes and responses to immune checkpoint blockade therapies \citep{Tumeh2014-wp,Chen2017-sf}. By leveraging multiplex imaging, researchers can map the spatial relationships between different cell types, revealing patterns of immune evasion or suppression that are critical for developing effective treatments.


To support and facilitate the analysis of high-level spatial architectures of tissues, recent studies have developed computational methods for unsupervised learning of spatial tissue structures from molecular tissue imaging data.
A widely adopted approach is to construct a graph of cellular interactions based on physical proximity, using it to cluster cells into distinct spatial regions \citep{Jackson2020-vg,Schurch2020-zh}. Building upon this, \cite{Kim2022-qu} developed an approach that integrates phenotypic information with the cell proximity graph to generate a new feature space encoding spatially aggregated phenotypic information, which is subsequently clustered using community detection algorithms such as Leiden \citep{Traag2019-ju}. 
Additionally, recent research has employed deep learning techniques, such as graph convolutional neural networks,  to combine spatial and phenotypic cellular data to model cell-cell interactions \citep{Fischer2021-ar} and discern cell neighborhood patterns \citep{Innocenti2021-pe}. \cite{Pham2020-kr} also proposed stLearn that utilizes deep learning for feature extraction from spatial molecular imaging data to obtain spatial clustering of cells.  While these approaches effectively incorporate spatial information into the identification of spatial tissue structures, most existing approaches rely on hard clustering methods, which presents significant limitations given the complex and often overlapping cellular communities within tissues. Hard clustering methods assign each cell to a single cluster without accounting for the fluid, intermingling nature of cellular communities. Consequently, they may miss subtle transitions between regions or nuanced tissue microenvironments that do not conform to rigid boundaries. Furthermore, they often lack the inferential benefits of statistical modeling, such as uncertainty quantification, both in identification of tissue structures and downstream association with clinical outcomes. 

\begin{figure*}[!t]%
\centering
\includegraphics[width=0.97\linewidth]{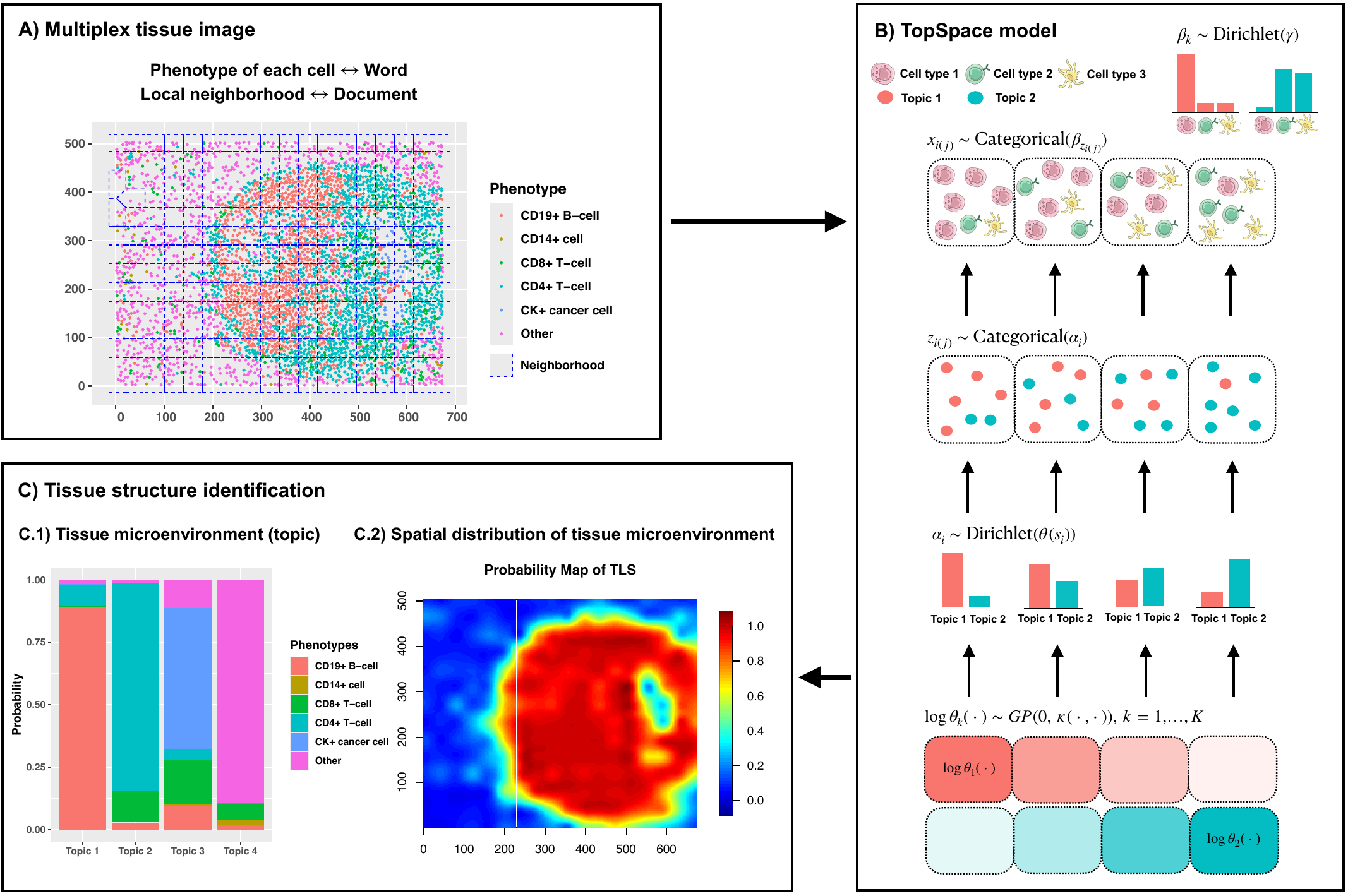}
\caption{Schematics of the \texttt{TopSpace} workflow. 
(A) \texttt{TopSpace} input data. \texttt{TopSpace} takes as input a multiplex image that includes individual cell phenotypes and pre-defined local neighborhoods. 
(B) A schematic representation of the \texttt{TopSpace} generative process. \texttt{TopSpace} leverages spatial GPs to account for spatial dependencies in the composition of tissue microenvironments (topics) across local neighborhoods. 
By analyzing co-occurrence of different cell types within these local neighborhoods while incorporating spatial information, \texttt{TopSpace} identifies latent tissue microenvironments (topics) characterized by unique distributions of cell types. 
(C) Spatial tissue structure identification using \texttt{TopSpace}. The results inferred by \texttt{TopSpace} are utilized to determine the spatial distribution of tissue microenvironments.
A high-resolution version of this figure is available in the supplementary material.}\label{fig:overviews}
\end{figure*}

To overcome the limited flexibility of hard clustering approaches in capturing complex biological tissue structures, we propose \texttt{TopSpace}, 
a novel Bayesian spatial topic model for unsupervised discovery of multicellular spatial tissue architectures in multiplex imaging data. Topic modeling is a machine learning technique commonly used in natural language processing to identify abstract themes within a collection of documents. Its mixed-membership nature makes it particularly suitable for flexibly modeling spatial tissue structures in multiplex imaging data. By applying topic modeling to this context, we examine cell types and local neighborhoods within a tissue section (Fig. \ref{fig:overviews}A)---analogous to words and documents in textual data---to identify biological ``topics,'' which are cellular microenvironments defined by the local distribution of cell types (Fig. \ref{fig:overviews}C. 1). 
Recently, \cite{Chen2020-sf} proposed a spatial topic model for multiplex imaging data that utilizes an adjacency-based spatial prior to account for spatial coherence among nearby cell neighborhoods. However, this spatial prior is limited to capturing spatial dependencies over pre-specified neighborhood adjacencies, which restricts its ability to model the intricate spatial dependence structures found in human tissues. We generalize this approach by employing spatial Gaussian processes (GPs), providing greater flexibility in capturing the complex and diverse spatial dependence structures observed within tissue samples. Moreover, we fully leverage the inferential benefits of Bayesian methods, such as uncertainty quantification and the selection of the number of topics through Bayesian model selection criteria.
Figure \ref{fig:overviews} illustrates the overall workflow of our \texttt{TopSpace}. The process begins with inputting a multiplex tissue image that includes individual cell phenotypes and pre-defined local neighborhoods (Fig. \ref{fig:overviews}A). \texttt{TopSpace} examines co-occurrence of different cell types within local neighborhoods while incorporating spatial context to identify biological topics, i.e., latent tissue microenvironments 
(Fig. \ref{fig:overviews}B). 
The inferential results from \texttt{TopSpace} are then used to determine the spatial distribution of tissue microenvironments (Fig. \ref{fig:overviews}C). 

We provide an efficient implementation of \texttt{TopSpace} using Markov chain Monte Carlo (MCMC), which enables accurate quantification of the uncertainty associated with the estimated microenvironments (topics), thereby fostering robust biological discoveries. 
Through rigorous simulations, we demonstrate that \texttt{TopSpace} is able to accurately recover latent topics and spatial clustering patterns in multiplex imaging data with varying levels of spatial dependencies. We also present a case study of non-small cell lung cancer (NSCLC) data, which aims to detect and quantify TLS using our \texttt{TopSpace} approach. 
Downstream survival analysis using the identified TLSs reveals a significant correlation between the presence of TLS and improved patient survival outcomes, demonstrating the practical applicability and impact of the proposed \texttt{TopSpace} in clinical research.

\section{Method}

In this section, we present our novel Bayesian spatial topic model, \texttt{TopSpace}, for the unsupervised discovery of spatial tissue architecture in multiplex imaging data. To ground our method, we first draw an analogy between topic modeling in natural language processing and the structure and inferential needs of multiplex imaging data, which will help clarify the motivation and approach of \texttt{TopSpace}.

\subsection{Correspondence between topic modeling and multiplex imaging data}

Topic modeling in natural language processing, such as probabilistic latent semantic analysis (PLSA, \citealp{hofmann1999probabilistic}) and latent Dirichlet allocation (LDA, \citealp{Blei2003-lq}), aims to identify latent themes in a collection of documents. In this framework, a ``document'' is composed of ``words'' which, through a probabilistic model, are grouped into``topics''---latent structures that reveal underlying themes. Analogously, when we work with multiplex imaging data, we can draw a similar correspondence (Fig. \ref{fig:overviews}A-B):
\begin{itemize}
    \item Document corresponds to a \emph{local neighborhood} within the tissue, which is defined by spatially proximal cells.
    \item Words correspond to \emph{observed cell phenotypes} within each neighborhood.
    \item Topics represent underlying \emph{multicellular microenvironments} that can be characterized by unique distributions of different cell types, embodying specific biological functions or interactions within the tissue.
\end{itemize}

In our setting, a tissue section is divided into small local neighborhoods, each of which corresponds to a document. 
There are multiple approaches available to construct local neighborhoods within a given multiplex image. An example is to select some anchor cells in the tissue section and construct neighborhoods centered around them. Another approach involves utilizing image partitioning techniques, such as Voronoi tessellation based on a specified set of points. In the experiments conducted in this paper, we generated grid points over the given tissue images and performed Voronoi tessellation based on these points. The grid points, which also define the spatial locations of the created neighborhoods, were arranged to ensure that cell-to-cell distances in each neighborhood remained below 30-40$\mu$m. This distance threshold was chosen because cell-cell interactions are typically local and weak beyond this range \citep{Mohammed2024-fr}.

Within each neighborhood, cell phenotypes correspond to biological ``words''. A topic, in this context, represents a specific microenvironment---a local distribution of cell types that characterizes interactions within the neighborhood. Our goal is to uncover these multicellular microenvironments, similar to inferring topics in LDA, but with spatial dependencies that capture the continuity of biological structures in the tissue. While LDA assumes independence among documents, the spatial nature of multiplex imaging data implies that nearby regions often share similar microenvironments. To address this, \texttt{TopSpace} incorporates spatial smoothing to encourage coherence among neighboring regions. 

\subsection{TopSpace model construction}

The proposed \texttt{TopSpace} model is a generalization of LDA to the analysis of multiplex imaging data, which inherently involves spatial information. While LDA assumes independence between documents, \texttt{TopSpace} incorporates the spatial dependency that is crucial for biological interpretation in tissue sections. Adjacent regions are more likely to share similar cellular microenvironments due to natural biological continuity, and our \texttt{TopSpace} employs spatial GPs to account for these spatial dependencies.

In a given multiplex image of tissue (Fig. \ref{fig:overviews}A), we assume that there are $H$ cell types that contribute to $K$ distinct cellular microenvironments (topics).
Let $M$ denote the number of local neighborhoods in the multiplex image and $N_i$ the number of cells in neighborhood $i \in \{1, \ldots, M\}$. For each cell $j \in \{1, \ldots, N_i\}$ in neighborhood $i$, we denote its observed phenotype by $x_{i(j)} \in \{1, \ldots, H\}$. 
Additionally, each cell is associated with a latent topic $z_{i(j)} \in \{1, \ldots, K\}$,  which governs the generation of cell type $x_{i(j)}$. In our \texttt{TopSpace} framework, we also observe the location of neighborhood $i$, which is denoted by $s_i$. 
The generative process of \texttt{TopSpace} is defined as follows:
\begin{enumerate}
\item For each topic $k$, draw a distribution over cell types from a Dirichlet distribution $\beta_k \sim \mbox{Dirichlet}(\gamma)$.
\item For each local neighborhood $i$, draw a distribution over topics from a Dirichlet distribution $\alpha_i \sim \mbox{Dirichlet}(\theta(s_i))$, where $\theta(s_i)= [\theta_1(s_i), \ldots, \theta_K(s_i)]^\top$ is the spatially varying Dirichlet hyperparameter that depends on the location $s_i$. 
\item For each cell $j$ in neighborhood $i$:
\begin{enumerate}
\item draw its latent topic assignment from $z_{i(j)} \sim \mbox{Categorical}(\alpha_i)$.
\item draw its observed cell type, conditioned on $z_{i(j)}$, from $x_{i(j)} \sim \mbox{Categorical}(\beta_{z_{i(j)}})$.
\end{enumerate}
\end{enumerate}

Note that our \texttt{TopSpace} model introduces a spatially varying Dirichlet hyperparameter $\theta(\cdot)$ for the Dirichlet prior on the neighborhood-specific topic distribution $\alpha_i$, which accounts for spatial dependencies across nearby cell neighborhoods. To directly model complex spatial dependence structures inherent in multiplex tissue imaging data, we assume a spatial GP prior over each element of $\theta(\cdot)$ subject to log transformation:
\begin{align}
\log \theta_k( \, \cdot \, ) \stackrel{ind.}{\sim} \mathcal{GP}(0, \kappa( \cdot , \cdot )), \ k = 1, \ldots, K, \label{eq:gp}
\end{align}
where $\kappa$ denotes the covariance kernel for GP.  The GP prior \eqref{eq:gp} imposes spatial smoothness on $\theta(\cdot)$, ensuring that the hyperparameters $\theta(s_{i_1})$ and $\theta(s_{i_2})$ are spatially coherent for nearby local neighborhoods $i_1$ and $i_2$. This spatial coherence propagates to the corresponding topic distributions $\alpha_{i_1}$ and $\alpha_{i_2}$, thereby inducing spatial dependencies in the distribution of topics across different tissue regions.  A schematic representation of the \texttt{TopSpace} generative process is presented in Figure \ref{fig:overviews}B.

Compared to existing spatial topic models, our proposed \texttt{TopSpace} offers notable advantages in modeling and interpreting spatial dependencies inherent in multiplex tissue image. 
First, unlike SLDA \citep{Wang2007-ny}, which jointly models both observed data and their spatial locations, \texttt{TopSpace} adopts a conditional modeling approach. Specifically, it models the region-specific topic distribution conditioned on spatial locations, which enhances its ability to directly capture and interpret spatial dependencies in the composition of tissue microenvironments. This conditional approach allows for more intuitive insights into how spatial arrangement influences tissue composition, making our \texttt{TopSpace} particularly compelling for elucidating spatial variations in cellular microenvironments in multiplex tissue imaging data.
Second, compared to spLDA \citep{Chen2020-sf}, \texttt{TopSpace} introduces greater flexibility in modeling complex spatial dependencies. spLDA relies on an adjacency-based similarity assumption, which enforces smoothness solely based on neighborhood adjacency. In contrast, \texttt{TopSpace} leverages nonparametric GPs to capture a wider range of spatial dependency patterns, accommodating the intricate and heterogeneous nature of biological tissues. This flexibility enables \texttt{TopSpace}  to better adapt to the diverse spatial structures observed within tissue samples (as we show in our simulations). 

In our \texttt{TopSpace}, the key inferential targets  are as follows:
\begin{itemize}
\item $\beta_k$: a tissue microenvironment (topic), characterized by a unique distribution of cell types (Fig. \ref{fig:overviews}C. 1). 
\item $z_{ij}$: the topic latently associated with each cell. 
\item $\alpha_i$: the probability distribution of topics across local neighborhoods. 
\end{itemize}
Once we obtain the inferential results for these parameters, they can be used for further downstream analyses to identify spatial tissue architectures (Fig. \ref{fig:overviews}C). First,
spatial clustering of multiplex imaging data can be achieved by clustering local neighborhoods based on the dominant topics with the highest posterior probabilities (see Fig. \ref{fig:nsclc}B). Moreover, the mixed-membership nature of our \texttt{TopSpace} allows for deeper quantification of tissue structures beyond simple spatial clustering. Our \texttt{TopSpace} model enables each local neighborhood to be associated with multiple topics with varying probabilities. 
By representing these variations as a spatial probability map of the tissue structures of interest (Fig. \ref{fig:overviews}C. 2)---whether focusing on a single topic or integrating several---we can quantify the tissue structure, for instance by computing quantitative imaging features. 

\subsection{Posterior inference}\label{sec:alg}

For posterior inference on our proposed \texttt{TopSpace}, we develop a computationally efficient MCMC algorithm to simulate the posterior distribution of \texttt{TopSpace}. Specifically, to improve scalability of our posterior computation procedure, we adopt a basis expansion of the GP priors using the eigendecomposition of the covariance kernel. With a sufficiently large set of eigenfunctions, the proposed \texttt{TopSpace} can be well approximated by a truncated linear combination of eigenfunctions, allowing us to develop an efficient Metropolis-Hastings within Gibbs sampling algorithm.  
In this algorithm, we use Gibbs sampling to update $\beta_k$ and $\theta_d$ and employ collapsed Gibbs sampling to update $z_{ij}$. The basis coefficients associated with the eigenfunction bases, which approximate the GP priors, are updated using the stochastic gradient Hamiltonian Monte Carlo proposed by \cite{Chen2014-lm}. Details of our MCMC sampling scheme are provided in the supplementary material.

\subsection{Selection of the number of topics $K$.} \label{sec:select-K} 
A key challenge in topic models is that the number of topics $K$ is generally unknown \emph{a priori}. This problem is especially prominent in the context of multiplex imaging data, where biological knowledge alone often cannot accurately specify the number of tissue microenvironments. 
To address this, we propose using the deviance information criterion (DIC, \citealp{Spiegelhalter2002-jb}), a Bayesian model selection criterion, to learn the number of topics $K$ directly from the data. 
Our approach follows the strategy introduced by \cite{Li2020-zm}, which provides an effective procedure to compute the DIC for latent variable models such as LDA. 
More details on computing DIC for \texttt{TopSpace} is provided in the supplementary material.

\section{Simulation studies}\label{sec:sim}

We assessed the performance of the proposed \texttt{TopSpace} using synthetic datasets designed to resemble our real data application (Section \ref{sec:reald}) with varying levels of spatial dependencies. We hypothesized that by accounting for the spatial dependencies inherent in the multiplex images, \texttt{TopSpace} would improve its estimation of latent topics and enhance the spatial clustering of tissue regions as spatial correlations increase.
We generated synthetic multiplex imaging datasets from the proposed TopSpace with $K = 3$ latent topics and varying degrees of spatial correlation. Specifically, we examined three simulation scenarios---denoted $ S_{low}, S_{med}$, and $S_{high}$---corresponding to low, medium, and high levels of  spatial correlation. Full details of the synthetic data generation procedure are provided in the supplementary material.
For each scenario, simulation results were summarized over 50 replications. 

We compared our \texttt{TopSpace} against two existing alternatives: (i) standard LDA \citep{Blei2003-lq}, which completely ignores  spatial information, and (ii) spLDA (\citealp{Chen2020-sf}), which accommodates spatial information through an adjacency-based spatial prior on $\theta_k$. 
spLDA involves a crucial tuning parameter $d_{ij}$ that governs the spatial coherence in the topic distribution between neighboring regions, which significantly affects its results. As \cite{Chen2020-sf} did not provide any guideline to select the value of this tuning parameter, we fitted spLDA with multiple values of $d_{ij}$, specifically, $d_{ij} = d \in \{0.025, 0.25, 2.5\}$. For the proposed \texttt{TopSpace}, we run the MCMC algorithm presented in Section \ref{sec:alg} for 20,000 iterations, of which the first 10,000 iterations were discarded as burn-in. 
We used the modified squared exponential kernel for the GP covariance kernel $\kappa$, fixing its two hyperparameters at $a = 0.01$ and $b = 1$. These hyperparameter values were correctly specified for the scenario $S_{med}$ but intentionally misspecified for $S_{low}$ and $S_{high}$.  Additional details on the choice of $\kappa$  are provided in the supplementary material.

\begin{figure*}[!t]%
\centering
\includegraphics[width=0.85\linewidth]{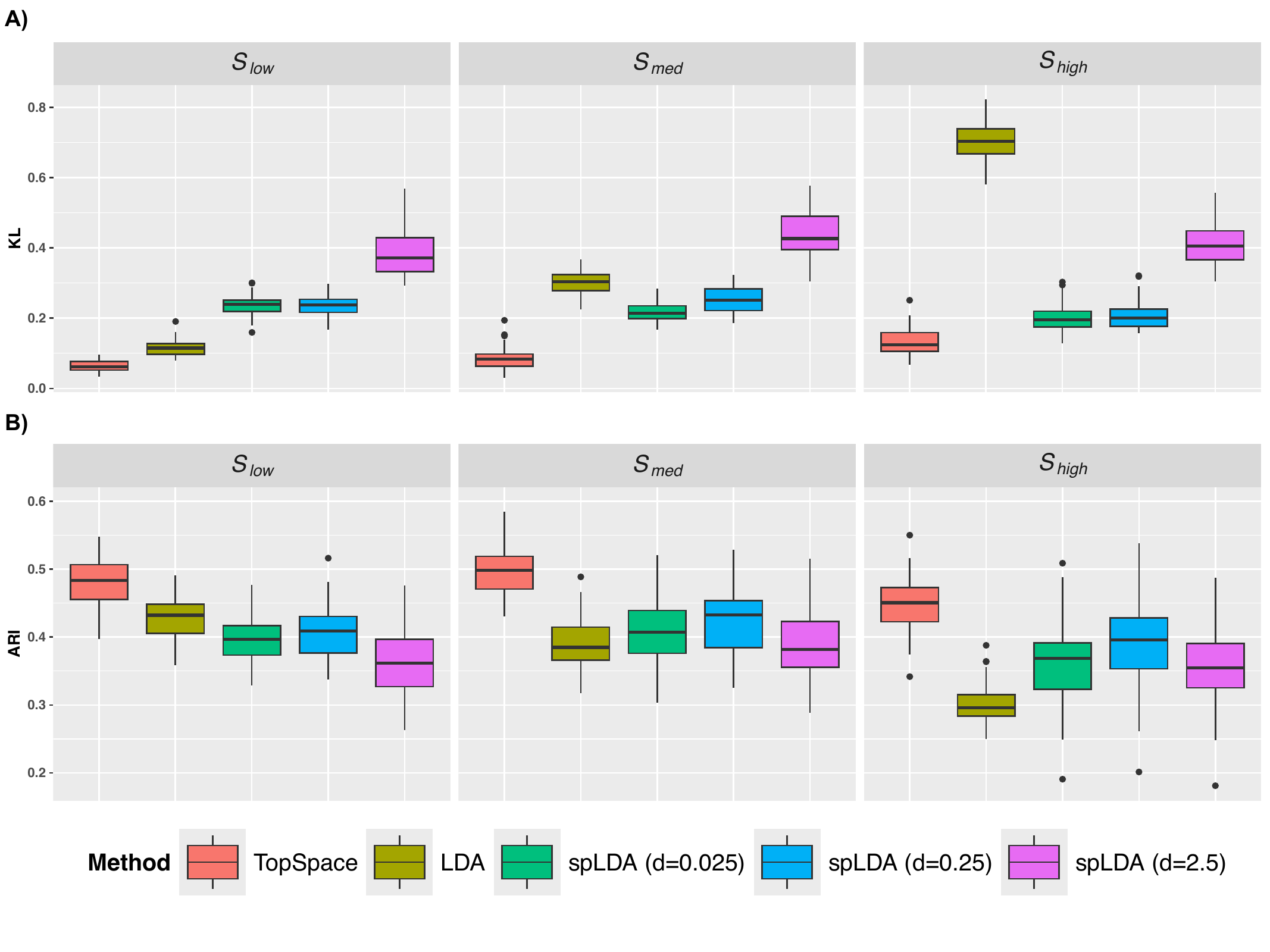}
\caption{Comparison of the performance of \texttt{TopSpace}, LDA, and spLDA 
in identifying latent topics and performing spatial clustering across simulation scenarios with different degrees of spatial dependencies. spLDA was fitted using various values for the tuning parameter $d_{ij}$, which regulates the strength of spatial coherence among adjacent neighborhoods. (A) Boxplots of the total KL divergence between true and estimated $\beta_k$'s, $\sum_{k=1}^K \mbox{KL}(\beta_k || \hat{\beta}_k)$, for each method, calculated across 50 replicates under varying degrees of spatial dependencies $S_{low}$, $S_{med}$ and $S_{high}$.  (B) Boxplots of ARIs comparing true and estimated spatial clustering for \texttt{TopSpace}, LDA, and spLDAs across 50 replicates  under different spatial dependency scenarios. A high-resolution version of this figure is available in the supplementary material.}\label{fig:sim}
\end{figure*}

\noindent\textbf{Latent topic estimation} 
As a measure of accuracy in estimating latent topics, we used the Kullback–Leibler (KL) divergence  between true and estimated $\beta_k$, summed over all $k$, i.e., $\sum_{k=1}^K \mbox{KL}(\beta_k || \hat{\beta}_k)$. Lower values of this measure indicate higher estimation accuracy. In Figure \ref{fig:sim}A, we present boxplots of the KL divergence for different methods across three different simulation scenarios. Our proposed \texttt{TopSpace} demonstrated superior performance in latent topic estimation compared to the competing methods, as evidenced by lower KL divergence scores. As spatial correlation levels increased, approaches incorporating spatial information (\texttt{TopSpace} and spLDA) showed lower KL divergence scores compared to the standard LDA which completely disregards spatial information. This demonstrates that incorporating spatial information into topic modeling enhances the accuracy of latent topic estimation, particularly for multiplex imaging data that inherently exhibit spatial correlations. Notably, \texttt{TopSpace} exhibited the best performance among spatially-informed approaches, highlighting its great flexibility in capturing complex spatial dependence structures.

\noindent\textbf{Spatial clustering} 
We also compared the spatial clustering performance of these methods. We computed the adjusted rand index (ARI) between true and estimated clustering membership, where the neighborhoods in the synthetic image are clustered based on the dominant topics inferred by each method. ARI values are ranging from 0 to 1, with 0 indicating no agreement between the two clusterings on any pair of points  and 1 indicating identical clusterings. A detailed formula for calculating the ARI is provided in the supplementary material. In Figure \ref{fig:sim}B, we provide the boxplots of the ARI for \texttt{TopSpace}, LDA, and spLDA across different levels of spatial dependency. Once again, \texttt{TopSpace} outperformed its competitors in spatial clustering across all the scenarios. Similar to the latent topic estimation, as spatial correlations increased, the topic models that account for spatial dependence structures (\texttt{TopSpace} and spLDA) demonstrated improved performances. Particularly,  with increasing level of spatial dependency, the difference in ARI between \texttt{TopSpace} and LDA became widened, with the mean difference of 0.052 for $S_{low}$, 0.108 for $S_{med}$ and 0.145 for $S_{high}$. 
This demonstrates the advantage of our \texttt{TopSpace} in effectively leveraging spatial information for clustering multiplex imaging data, especially in cases where significant spatial correlations exist. Moreover, it should be highlighted that despite the hyperparameters for the GP covariance kernel being misspecified for the simulations $S_{low}$ and $S_{high}$, \texttt{TopSpace} still outperformed its competitors in both latent topic estimation and spatial clustering,  demonstrating its robustness to hyperparameter tuning.

We also evaluated our proposed method for choosing the number of topics $K$ using the DIC. We considered different numbers of topics $K \in \{2, 3, 4, 5\}$ and selected the optimal number of topics based on the DIC. Across the simulation scenarios, \texttt{TopSpace} reliably identified the correct number of topics with high accuracy: 96\% for $S_{low}$, 94\% for $S_{med}$, and 98\% for $S_{high}$. This highlights the effectiveness of DIC for choosing $K$ in our \texttt{TopSpace} framework.

\section{TLS identification in NSCLC data}\label{sec:reald}

\begin{figure*}[!t]%
\centering
\includegraphics[width=\linewidth]{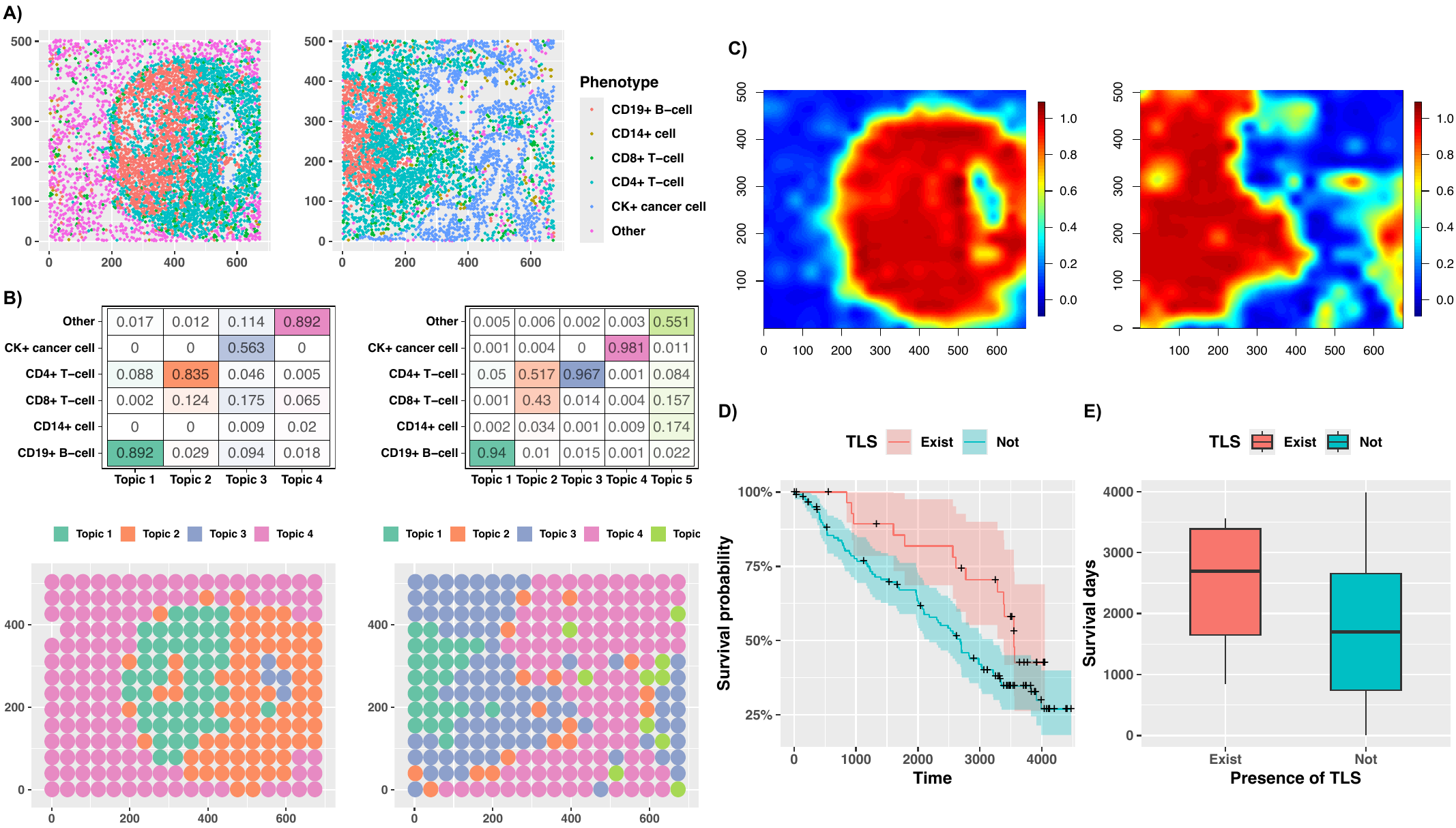}
\caption{Analysis of the non-small cell lung cancer dataset collected from 153 patients. (A) Representative images displaying 6 cell types, with colors specified in the legend. (B) Estimated topics (per-topic cell type distributions) and spatial clustering based on dominant topics, achieved through the application of \texttt{TopSpace} to the representative images. Selection of the number of topics was guided by the DIC, choosing four topics for the first image and five for the second. Topics are arranged to highlight specific microenvironments, with Topic 1 representing the B-cell zone and Topic 2 (as well as Topic 3 for the second image) representing the T-cell zone. (C) The spatial probability map of TLS for each image, created by combining probabilities of Topics 1 and 2 (and Topic 3 for the second image) across local neighborhoods. Areas in red and blue depict high and low probabilities of TLS, respectively. (D) Kaplan-Meier survival curves for patients with and without TLS. (E) Boxplots comparing survival days between patients with and without TLS, using data from uncensored patients only; significant difference indicated by the p-value of 0.007 from the Wilcoxon test. A high-resolution version of this figure is available in the supplementary material.}\label{fig:nsclc}
\end{figure*}

We demonstrate the utility of the proposed \texttt{TopSpace} using a non-small cell lung cancer (NSCLC) dataset collected through multiplex immunohistochemistry \citep{Johnson2021-fp}. The NSCLC dataset comprises immunohistochemistry images of tissue sections from $153$ NSCLC patients. Tissue sections were stained with markers that include CD19, CD8, CD3, CD14, major histocompatibility complex II, cytokeratin, and DAPI. Following image processing, which involves cell segmentation, phenotyping, and batch correction, each image was converted into a data matrix, where rows correspond to individual cells and columns correspond to x- and y-coordinates of each cell on the image, individual marker expression, and cell phenotypes. There are six cell phenotypes identified in this dataset: $\text{CD19}^+$ B-cells, $\text{CD14}^+$ cells, $\text{CD8}^+$ T-cells, $\text{CD4}^+$ T-cells, $\text{CK}^+$ cancer cells, and Others.  For each patient in the NSCLC dataset, three to five images were captured from small regions of the tissue sample. To account for sparsity in some images, we selected the image with the highest cell count to represent each subject. Figure \ref{fig:nsclc}A shows two representative images from two selected patients. Further details on the NSCLC dataset are available in \cite{Johnson2021-fp}. 

In our study of the NSCLC dataset, we focused on TLS identification using \texttt{TopSpace}. TLSs are ectopic lymphoid formation that emerges under persistent inflammatory conditions, such as tumors, and primarily consist of aggregates of B-cells and T-cells at various organizational stages \citep{Chen2022-tb}. These structures are crucial for the optimal development and coordination of the adaptive immune response, delivering an augmented effect on the tumor microenvironment. It has been observed that the presence of TLS affects the efficacy of cancer immunotherapies and is strongly associated with improved survival and clinical outcomes in patients with various cancers, including NSCLC \citep{Tang2020-rw}.

\ech

TopSace was applied to each patient image. 
For each image, we considered various numbers of topics $K \in \{2, 3, 4, 5\}$, selecting the optimal one based on the DIC. Considering the organized structure of TLSs with segregated B- and T-cell zones \citep{Chen2022-tb}, 
we identified TLS by searching for topics dominated by B-cells and T-cells, respectively. A topic was classified as representing the B-cell zone if the probability of $\text{CD19}^+$ B-cell exceeded 0.5, and the T-cell zone if the probability of $\text{CD4}^+$ and $\text{CD8}^+$ T-cells exceeded 0.5. Figure \ref{fig:nsclc}B displays the estimated topics and spatial clustering for each representative image. These images demonstrated topics representative of both the B-cell zone (Topic 1) and T-cell zone (Topic 2 for the first image and Topics 2 \& 3 for the second image). The co-existence of both topics within an image indicated the presence of TLS.
For images where TLS was detected, we also visualized the spatial probability map of TLS (Fig. \ref{fig:nsclc}C) by combining  probabilities of the topics representing the B-cell and T-cell zones. These maps, highlighting areas in red and blue for high and low TLS probabilities, respectively, provide a more informative view than simple spatial clustering and allows for further quantification of extant of TLS, such as calculating quantitative imaging features. 

We examined the association between the presence of TLS, identified using \texttt{TopSpace}, and patient survival. TLSs were detected in 30 out of 154 patient images, and we performed survival analysis based on these findings. Figure \ref{fig:nsclc}D presents the Kaplan-Mier survival curves for patients with and without TLS, illustrating improved survival probabilities for patients with TLS. We also compared the survival days of patients, without censoring, between groups with and without TLS. Figure \ref{fig:nsclc}E displays boxplots comparing these survival days, where the Wilcoxon test returned a significant p-value of $0.007$, indicating a meaningful difference in survival. Additionally, we investigated the relationship between patient survival and quantitative imaging 
features derived from the TLS spatial probability map (Fig. \ref{fig:nsclc}C), with detailed results provided in the supplementary material.

\section{Discussion}

We have developed a novel Bayesian spatial topic model, \texttt{TopSpace}, to uncover the multicellular spatial architecture of tissues and identify characteristic cellular microenvironments in multiplex tissue images. The proposed \texttt{TopSpace} leverages spatial GPs to incorporate spatial information into topic modeling, flexibly accounting for spatial dependencies within nearby local neighborhoods and effectively adapting to diverse spatial patterns of tissue organization observed in multiplex tissue images. Simulation studies show that accounting for spatial dependencies in spatially structured tissue images improves the effectiveness and accuracy of topic estimation and spatial clustering. In a case study using the NSCLC dataset, \texttt{TopSpace} successfully identified TLSs, which are strongly linked to improved survival and better responses to cancer immunotherapies in cancer patients. Going beyond spatial clustering, \texttt{TopSpace} also generated spatial probability maps of TLS, which depict the likelihood (and extant) of the presence of TLS across tissue regions and serve as a powerful tool for TLS quantification. Survival analysis with identified TLS showed that the presence of TLS is associated with higher survival rates in NSCLC patients, demonstrating the utility of \texttt{TopSpace} in clinical research.


Our MCMC implementation of \texttt{TopSpace} takes an average of 15.5 minutes per image for analyzing the NSCLC dataset on an Intel i9-9880H 2.3GHz CPU. In this application, we ran four MCMC chains sequentially for each value of $K \in \{2, 3, 4, 5\}$; however, these computations can be parallelized to reduce the computational time. Our \texttt{TopSpace} R package is available at \url{https://github.com/junsoukchoi/TopSpace.git}.

Despite its strengths, \texttt{TopSpace} has certain limitations that present opportunities for future research. One limitation is that it is currently designed for the analysis of a single image at a time. When applied to multiple images, each image must be modeled separately, preventing the model from leveraging shared structures across images. A promising direction for future research is to extend \texttt{TopSpace} to a joint modeling framework for multiple images, allowing the borrowing of strength across images and improving the robustness and generalizability of the inferred spatial microenvironments. 
Another limitation is that \texttt{TopSpace} does not incorporate  patient-level information, such as tumor stage or response to immunotherapy, which is often available in multiplex imaging data. 
A future version of \texttt{TopSpace} could integrate patient-level predictors to guide the discovery of biologically significant tissue structures, potentially revealing clinically relevant associations between spatial tissue architectures and patient outcomes.  

Beyond its application to multiplex imaging data, the general framework of \texttt{TopSpace} has broader applicability. The methodology is well-suited for other spatial omics datasets, such as spatial transcriptomics, where observations are categorized into discrete types, such as cell types or gene expression clusters. The flexibility of the Bayesian spatial topic modeling approach suggests that \texttt{TopSpace} could be adapted to analyze diverse forms of spatially structured biological data, further expanding its utility in biomedical research.


\section{Key Points}

\begin{itemize}
\item We propose TopSpace, a novel Bayesian spatial topic model for  unsupervised discovery of multicellular spatial tissue structures in multiplex imaging data. 
\item TopSpace leverages inferential benefits of Bayesian modeling, including uncertainty quantification and data-driven selection of the optimal number of tissue microenvironments. 
\item Applied to non-small cell lung cancer data, TopSpace successfully identifies tertiary lymphoid structures and characterizes their spatial distributions within lung cancer multiplex images, which strongly correlate with patient survival outcomes---highlighting its biomedical relevance.
\item Extensive simulation studies show that TopSpace accurately recovers latent tissue microenvironments and spatial clustering patterns, outperforming existing methods across a range of spatial dependency settings.
\item We provide a user-friendly R package, TopSpace, to support reproducibility and facilitate broader adoption of our proposed method.
\end{itemize}



\section{Funding}

V.B. was supported by NIH grants R01-CA160736, R01CA244845-01A1, and P30 CA46592. J.K. and J.C. were partially supported by NIH grant R01MH105561 and NSF grant IIS2123777. 

\section{Author Biographies}

Junsouk Choi, PhD, is an Assistant Professor in the Department of Statistics at Korea University, South Korea. 
His research interests center on  developing Bayesian and  machine learning methods for high-dimensional, complex-structured data in biomedical fields, including imaging, genomics, and related areas.

\vspace{0.1in}

\noindent Jian Kang, PhD, is a Professor and Associate Chair for Research in the Department of Biostatistics at the University of Michigan, Ann Arbor. His primary research interests are in developing statistical methods for large-scale complex biomedical data with application in precision medicine, imaging, epidemiology and genetics.

\vspace{0.1in}

\noindent Veerabhadran Baladandayuthapani, PhD, is Professor and Chair of the Department of Biostatistics at the University of Michigan, Ann Arbor, and also serves as Associate Director of the Center for Cancer Biostatistics.
His research interests are mainly in developing Bayesian probabilistic models and machine learning methods to assist in medical and health sciences.

\bibliographystyle{unsrt}
\bibliography{reference}

\end{document}